# Sensitivity of Standard Library Cells to Optical Fault Injection Attacks in IHP 250 nm Technology

Dmytro Petryk[1], Zoya Dyka[1] and Peter Langendörfer[1,2]

[1]IHP – Leibniz-Institut für innovative Mikroelektronik
Frankfurt (Oder), Germany

[2]BTU Cottbus-Senftenberg
Cottbus, Germany

*Abstract*—**The IoT consists of a lot of devices such as embedded systems, wireless sensor nodes (WSNs), control systems, etc. It is essential for some of these devices to protect information that they process and transmit. The issue is that an adversary may steal these devices to gain a physical access to the device. There is a variety of ways that allows to reveal cryptographic keys. One of them are optical Fault Injection attacks. We performed successful optical Fault Injections into different type of gates, in particular INV, NAND, NOR, FF. In our work we concentrate on the selection of the parameters configured by an attacker and their influence on the success of the Fault Injections.**

## I. Introduction (Motivation)

The Internet of things (IoT) is becoming increasingly essential in our lives. One of the reasons is technology miniaturization. It allows to reach less power consumption and place more transistors per chip area that leads to more cost effective manufacturing. These advantages are important criteria to deploy WSNs in a large volume. However, with technology miniaturization semi-conductor devices suffer more from radiation and other sources of faults, e.g. fluctuation of temperature, frequency, voltage, light influence, etc. Hence, optical (laser) Fault Injection (FI) attacks become more likely. Optical FI attack belongs to the semi-invasive type of attack [1]. This type assumes that an attacker has to gain a physical access not only to the device itself but also to its internal structure. Hence, it usually requires some preliminary preparations of the attacked device, e.g. decapsulation of an attacked chip. Details about all other types can be found in [1]. The goal of FI attack is to induce an error that may switch the device into unintended operation mode. Exploiting this operation state of a device and observing its output the sensitive data may leak.

Scaling technologies however make it harder to influence a single transistor. This is due to the physical sizes of a transistor and limited minimal spot size of a laser beam that subsequently covers the area of many gates.

In essence laser based attacks are possible due to the photoelectric effect. This effect is observed in different materials such as semi-conductors, dielectrics and metals. This effect is based on the interaction of silicon with the electromagnetic radiation in a visible spectrum, i.e. light. When the laser emits its beam that passes through the silicon the electrons in it absorb energy from photons and become free, i.e. electrons move to the conductive band. This leads to a measurable increase in the current. The increase in the current is transient as it vanishes when all charges are recombined. More details about the internal photoelectric effect in semi-conductors can be found in [2]. Usually after the increase in current devices begin to work as specified by the manufacturer.

Hence, if accurate timing and precise spatial location of a laser beam are ensured optical FI attacks can influence the state of a selected transistor i.e. the transistor may switch from a high resistance state (closed) to a low resistance state (open). Successful FI attack may lead to different unexpected faulty states of an attacked device. Depending on the effect of optical FI attack, faults are classified into following models:

- Bit-set: the state of the attacked cell changed: '0' → '1'.
- Bit-reset: the state of the attacked cell changed: '1' → '0'.
- Bit-flip: the state of the attacked cell changed into opposite logical state.
- Stuck-at: the change of the attacked cell state is no more possible.

The success of optical FI attacks depends on many parameters which have to be considered. Only an appropriate set of parameters may lead to reproducible faults. These parameters are: wavelength of the selected laser, spot size of the laser beam, chip position (X, Y, Z), timing, pulse duration and intensity of the laser beam

## II. Attacked chip

In order to assess the sensitivity of standard library cells of the IHP 250 nm technology to optical FI attacks we experimented with an old IHP chip "Libval025" produced without metal fillers. Its internal structure is clearly visible through a microscope. The chip was designed to measure signal propagation delays through different type of gates. It contains: Inverter (INV), NAND, NOR and Flip-flop (FF) gates. Each chain contains 4 lines with different loads. In total Libval025 has 16 lines. It is operated by 10 pins: input (in), output (out), reset, 3 power lines (vdd_c, vdd_p, GND) and 4 selection line pins (out_sel0, out_sel1, out_sel2, out_sel3). The structural scheme of Libval025 is shown in **Fig. 1**.

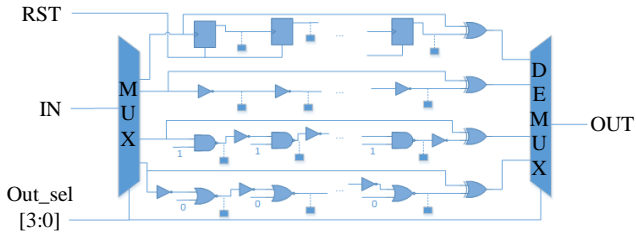

**Fig. 1.** Structural scheme of Libval025.

Only 1 of 16 lines, i.e. the one selected by the MUX, is active in the attacked Libval025 chip. Selection of the line is achieved by sending signals ('0' or '1') to selection pins. These signals go to the input of the multiplexer and it directs a signal from the input pin to explicitly selected line only. All other lines receive an input '0' and do not switch during operation. Each selected line of FF, NAND, NOR has 200 cells connected in series. Each selected line of INV has 600 cells connected in series. The 3 Libval025 chips attacked are placed on a PCB are shown in **Fig. 2**.

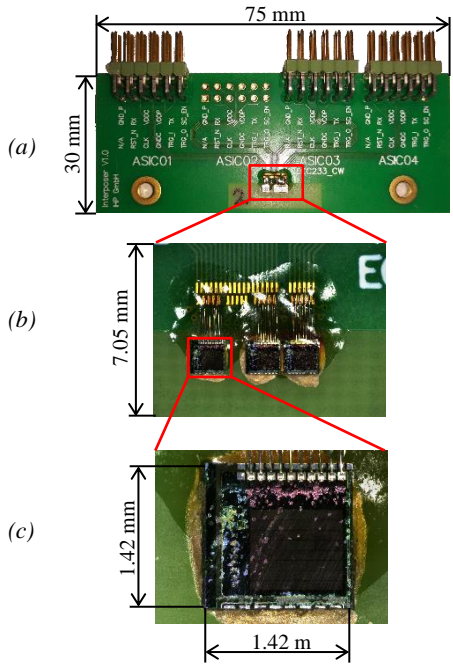

**Fig. 2.** The attacked chips:
*(a)* – PCB with 3 Libval025 chips; *(b)* – 3 Libval025 chips, zoomed in; *(c)* – single IHP Libval025 chip, zoomed in.

After bonding, wires of the chips were covered with an epoxy in order to protect them from mechanical damage.

## III. FAULT INJECTION SETUP

In order to perform an attack on Libval025 we use a setup that consists of: a Diode Laser Station (DLS), a VC glitcher, PC with Riscure Inspector software [4], a stable power supply and an oscilloscope. The structural view of our FI setup is shown in **Fig. 3**.

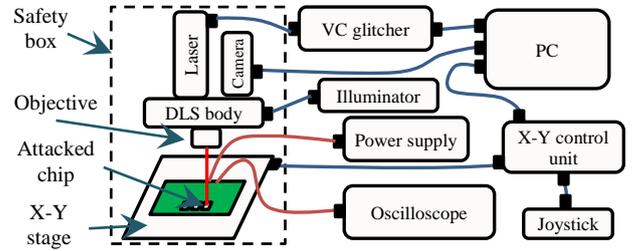

**Fig. 3.** Optical Fault Injection setup as a part of the laboratory equipment at IHP.

The basis of the FI setup is the 1st generation Riscure DLS [3] that is placed in a safety box to protect the user from the possibly harmful reflections of the laser beam. The Diode Laser Station is shown in **Fig. 4**.

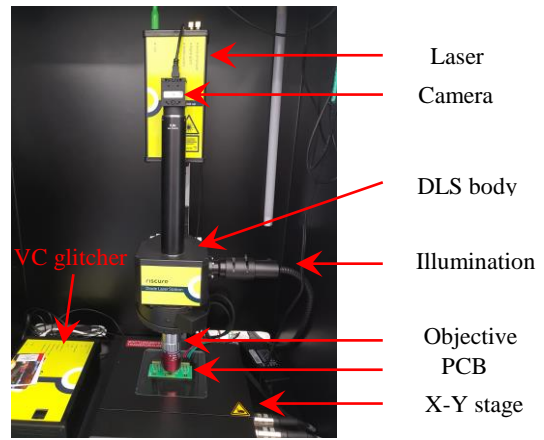

**Fig. 4.** Diode Laser Station at the IHP.

The DLS consists of: laser source, camera, illuminator, DLS body, optical system, X-Y positioning stage. DLS has following specifications [3]:

- Two multimode laser sources with maximum power of 14 W, red laser (808 nm); 20 W, infrared laser (1064 nm);
- pulse duration in the range of 20 ns – 100 μs;
- elliptical spot sizes 60×14 μm$^2$, 15×3.5 μm$^2$ or 6×1.5 μm$^2$;
- magnification objectives: 5×, 20×, 50×;
- Filter: 0.1 %, 1 %, 10 %;
- X-Y table with 3 μm accuracy and 0.05 μm [3] step size.

The PC with Inspector software [4] allows to create a FI program where the user sets parameters of the laser source (pulse duration and output power) and X-Y stage, i.e. an area of automatic scan. The PC is connected with the X-Y stage, microscope camera and a VC glitcher separately via USBs. The VC glitcher [5] is a control unit that allows to control the optical pulses of the DLS. It firstly loads and processes an FI program received from the PC and then sends commands to a laser source via SMA cables. The X-Y stage is controlled by the Inspector software. It allows to move an attacked chip and perform an automatic scan by setting an area and a step size. The X-Y stage is produced by Märzhäuser Wetzlar GmbH & Co [6].

## IV. EXPERIMENTS AND RESULTS

We attacked the Libval025 through the front-side. This is due to the fact that the structure of the chip is visible through the microscope camera, easily accessible and requires no preliminary preparations. This choice led us to the use of a multimode red laser (808 nm) as it has higher photon energy despite lower penetration depth/higher absorption coefficient. Our first goal was to define a set of parameters which ensures successful FI. Due to the fact that we knew nothing about the sensitivity of cells to optical FI in IHP 250 nm technology we started with the lowest energy and minimal pulse duration of the laser source to avoid damage of the Libval025. In case of a failed attack, i.e. no injected fault(s), we increased the power or pulse duration gradually. We selected the INV line for the first laser attack. We performed an attack with the following parameters: 1 % power, 20 ns pulse duration, 0.1 % filter, 50× objective and 1 µm step size for X and Y axis. The size of the laser spot when using the 50× objective is 6 µm × 1.5 µm corresponding to the DLS data sheet [3].

The area of the scan was limited by one cell in a selected line, i.e. one inverter (INV cell). We observed that first transient faults were injected with the following parameters: 80 % power, 100 ns pulse duration, 10 % filter, 50× objective and 1 µm step size for X and Y axis. **Fig. 5** shows the oscilloscope waveforms of input and output of the Libval025 chip measured during scan.

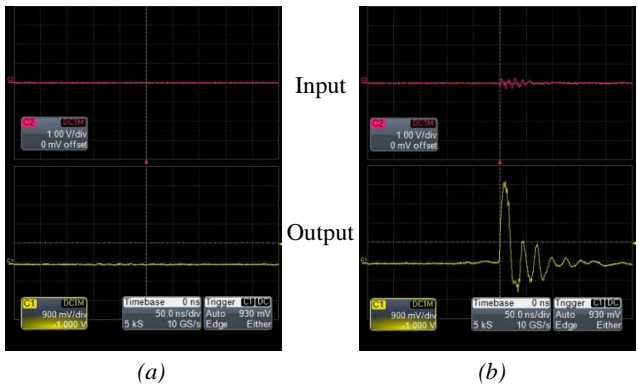

**Fig. 5.** Oscilloscope waveforms of input and output of our Libval025 chip measured during: *(a)* – normal operation; *(b)* – successful FI.

Then we attacked other cells in the same selected line with the same set of parameters. We observed successful and repeatable faults. Afterwards we repeated similar experiments with NAND, NOR, FF lines, also with different loads. We are able to inject transient repeatable faults with a similar but not the same FI parameters (power of the laser beam, pulse duration, etc). To prove our observations we attacked two other Libval025 chips. The results were similar with a slight difference in parameters, e.g. higher/lower power or longer/shorter pulse duration. Hence, we can conclude that transient faults may be successfully and repeatedly induced in INV, NAND, NOR, FF standard cells in a IHP 250 nm technology with the following set of parameters: 85 % power, 100 ns pulse duration, 10 % filter, 50× objective and 1 µm step size for X and Y axis. According to our observations and results obtained we assume that we opened an NMOS transistor in the INV, NAND, NOR gates. Please note that one input of each NAND gate is always connected to the power supply line (vdd), i.e. to the logical '1' (see **Fig. 1**) and one input of each NOR gate always is connected to GND i.e. to the logical '0'. Due to the logical state of the gates attacked the injected faults are classified as bit-resets. However for the FF gates we observed bit-sets only. Despite successfully injected transient faults in all types of gates we continued to increase the power in order to assess how this may influence the behavior of the gates. Damage of the internal structures of the gates was observed with the following parameters: 24 % power, 50 µs pulse duration, no filter, 50× objective. After this damage the line takes a value ('0' or '1') that can`t be changed by changing input values or powering the chip on/off. This effect was observed for all gates in Libval025.

TABLE I summarizes the results of attacks on Libval025 chips.

TABLE I. RESULTS OF ATTACKS ON LIBVAL025 CHIPS

| Fault | FF | INV | NAND | NOR |
|---|---|---|---|---|
| Stuck-at fault | ✓ | ✓ | ✓ | ✓ |
| Bit-set | ✓ | × | × | × |
| Bit-reset | × | ✓ | ✓ | ✓ |

## V. CONCLUSION

This work presents that standard library cells in IHP 250 nm technology are sensitive to optical FI attacks. Hence, performing optical FI attacks allows to retrieve or modify the data processed by devices based on this technology. We injected faults precisely despite the fact that the area of laser beam's spot is comparable with the area of one inverter in the attacked technology. However the current requirement of IHP technology is to add metal fillers. This can decrease the success of FI attacks significantly. Metal fillers can be used as a kind of cheap countermeasures if the gate area, that is sensitive to optical faults, is covered. In our future work we plan to determine the sensitive area of different gates with the goal to propose a metal fillers placement methodology that can be a part of an (automated) chip design flow.

ACKNOWLEDGMENT

This project has received funding from the European Union's Horizon 2020 research and innovation program under the Marie Skłodowska-Curie grant agreement No 722325.